\documentclass[apjl]{emulateapj}

\def\gtorder{\mathrel{\raise.3ex\hbox{$>$}\mkern-14mu
             \lower0.6ex\hbox{$\sim$}}}
\def\ltorder{\mathrel{\raise.3ex\hbox{$<$}\mkern-14mu
             \lower0.6ex\hbox{$\sim$}}}

\shorttitle{Migration and Survival of Terrestrial Planets}
\shortauthors{Mandell \& Sigurdsson}

\begin{document}
\title{Survival of Terrestrial Planets in the
Presence of Giant Planet Migration} 
\author{Avi M. Mandell \& Steinn Sigurdsson\footnotemark[1]
}
\affil{}

\begin{abstract}
The presence of ``Hot Jupiters'', Jovian-mass planets with very short
orbital periods orbiting nearby main sequence stars, has been proposed
to be primarily due to the orbital migration of planets formed in
orbits initially much further from the parent star.  This migration
affects the evolution of inner
terrestrial planets in these systems. Previous analyses have
assumed that no terrestrial planets survive after migration has
occurred.  We present numerical simulations showing that a significant
fraction of terrestrial planets could survive the migration process,
and possibly return to near circular orbits relatively close to
their original positions.  A fraction of the final orbits are in the
Habitable Zone, suggesting that planetary systems with close-in giant
planets are viable targets for searches for Earth-like habitable
planets around other stars.
\end{abstract}   


\section{}

\footnotetext[1]{ 
525 Davey Laboratory, Department of Astronomy \& Astrophysics, 
Pennsylvania State University, University Park, PA 16803, USA; 
mandell\@astro.psu.edu, steinn\@astro.psu.edu}
Over 100 extra-solar planets have been observed \citep{sch02}, and
more discoveries are announced each month.  The existence of
giant planets at very small orbital radii from their host stars
was one of the major surprises that has emerged from these detections
\citep{may95}.  As long-duration studies increase their sensitivity
to long-period planets, there remains a significant
accumulation of planets in circular orbits at distances $\ltorder $ 0.25 AU.
Though current detection techniques are
biased towards massive planets in close orbits, this pile-up at small
distances appears to be genuine.

Current theories postulate several different mechanisms to explain the
phenomena of giant planets with small orbital radii.  It is generally
believed that these planets formed at larger distances either in the
conventional model of coalescence of planetesimals to form a core onto
which gas accreted \citep{pol96} or by direct gravitational collapse
due to instabilities in the disk \citep{bos01,may02} and then migrated
inwards due to one, or more, of the following processes: ``early
migration'' involving tidal interaction with the evolving
circumstellar disk \citep{lin96}, or ``late migration'' involving
planetesimal scattering \citep{mur98} and/or gravitational scattering
with other planets \citep{wei96,ras96}.  Gravitational scattering alone
cannot account for the shortest period Jovian-type planets in circular
orbits, so dissipative migration must occur in some
systems.  Models suggest that migration processes are an
inherent result of disk-planet interactions, and time scales for
the duration of migration, after formation of the most massive planet,
range from $10^{5}$ to $10^{6}$ years, decreasing as initial disk mass
increases and the initial planet formation distance decreases
\citep{tri02}.  
A number of
studies have explored the dynamical stability of terrestrial-sized
bodies in systems with close-in giant planets. \citet{men03} performed
an exhaustive examination of all detected systems, concluding that at
least 25\% of systems permit stable terrestrial planets in the
Habitable Zone \citep{kas93} of their star, and considerably more
allow a low-mass planet outside of the conventional habitable
parameter space.  However, the dynamical effects of migration on
existing terrestrial planets have not been examined in depth.

Terrestrial planet formation is thought to occur in a ``runaway
growth'' scenario, where the larger gravitational cross-section of the
largest planetesimals increases the rate of interactions and
facilitates the growth of several large cores \citep{kor01, cha01}.
Current dynamical models predict that Mars-sized objects form in the
inner system on a time scale of order $10^5$ years and terrestrial
planets are fully formed within 30 million years, but recent
radioactive dating suggests that our inner Solar System may have
formed in as little as 10 million years \citep{jac03, yin02, kle02}.
Similarly, time scales for giant planet formation due to core accretion
range from $10^6$ to $10^7$ years \citep{pol96}, with Type II
migration beginning after the planet has reached $\sim10-30
M_{\earth}$ \citep{lin93}; therefore the initial onset of giant planet
migration may be relatively slow ($10^{6}$ years or greater).  This
would suggest that substantial terrestrial-sized planetesimals, or
``planetary embryos'', may already be present in the inner system when
giant planet migration begins.  Giant planet migration also has
profound implications for the evolution of the remaining circumstellar
material.  \citet{arm03} examined the effect of migration on an
initial gas disk, and found that late migration would cause an inward
flow of dust-depleted gas which would suppress late terrestrial planet
formation.  However, the author did not consider the dynamical effects
on any planetesimals, either within the migration radius or outside of
it.  \citet{the02} examined the effect that several different giant
planet systems with unusual properties (high eccentricity, multiple
planets) would have on terrestrial planet formation, but did not
address migration at all.

To investigate the potential effects of giant planet migration on a
nascent system of terrestrial planets, we have constructed a set of
dynamical planetary system models using a hybrid symplectic integrator
modified to include artificial giant planet migration.  The
current simulations assume that a giant planet is relatively close to
its final mass before it begins to migrate, and the terrestrial
planets are at a substantial fraction of their final mass.  Since the
details of planetary evolution in systems which experience Jovian
migration are currently largely unconstrained, the assumptions
included in these simulations may need to be modified for future work,
especially if the time scales for terrestrial and giant planet
formation prove to be radically different.  However, examining the
general dynamical evolution of this simple model should help to define
the basic parameters of the problem and illuminate interesting areas
for future study.

Numerical simulations were performed using a modified version of the
publicly-available hybrid symplectic integrator package MERCURY by
Chambers \& Migliorini \citep{cha97,cha99}.  To examine the effects of
migration, we modified the integrator to accommodate a
secular decrease in the semi-major axis of a giant planet. 
With each time step, the coordinates were adjusted so
that the planet would move in towards the star at the specified rate.
Since the model migration is included by fiat, not through
internal physically motivated processes, the system is no longer
energy conserving, but the forced change in orbital parameters over
each time step for a migration time scale of $10^6$ is small, and
integration parameters set for a stationary system provide accurate
integration of the planets' interactions.
The mechanism for migration is not important in these simulations
since we are purely investigating the effects of a migrating Jupiter
on terrestrial planets at different times.  In future simulations, a
more physical mechanism for migration may be introduced to more fully
analyze effects on the giant planet itself.

The possibility of Type I migration of the terrestrial bodies
due to disk interactions (as described by \citet{war97} and
\citet{kor01}) was ignored for the main body of simulations for
several reasons.  First, Type I migration time scales are rapid (on the
order of $10^5(M/M_{\earth})^{-1}$ where M is the mass of the planet
\citep{war97}), so the separations between the giant planet and any
terrestrial planets interior to the giant planet (except for objects
smaller than $0.1M_{\earth}$) would increase and very few
terrestrial-Jovian interactions would occur.  Any interactions that
would occur would be governed by the same stochastic processes as in
the non-migrating case since these terrestrial planets would be
migrating slower than the Jovian-type planet.  Several test runs with
Type I migration included confirmed these assumptions; only the
smallest planets interacted with the migrating Jupiter, and for these
the survival statistics were similar to the simulations without Type I
migration.  Second, recent results \citep{nel03,lau03} suggest that
for bodies smaller than $\sim10M_{\earth}$ MHD turbulence will cause the
motion due to torques from the disk to occur in a random walk, and
over long time periods smaller bodies will tend to maintain their
average position.

For the investigation of the dynamical evolution of terrestrial-sized
bodies, several different configurations of bodies with different
masses and radii were used.  As a starting point
an inner system was
set up consisting of four bodies with masses and initial orbital
parameters identical to our current inner Solar System, since we have
reliable data on that configuration.  A planet
identical to Jupiter was placed at 5.2 AU and allowed to migrate
inwards over three different time scales: $5\times10^{5}$,
$1\times10^{6}$, and $2\times10^{6}$ years.  In order to avoid
uncertainties in the variability of migration rate due to disk
properties, a constant migration rate was used to compare different
migration time scales, and additional simulations were performed to
compare constant migration to a more realistic migration model taken
from the literature \citep{tri02,lin86}.  One hundred integrations
were performed for each migration time scale using a constant migration
rate, and another one hundred integrations were performed using a
variable migration rate over $1\times10^{6}$ years.  The initial
anomalies for the four terrestrial planets were randomly selected for
each integration.  In addition to the above model, simulations were
run with three different configurations of non-Solar planetary systems, taken from the
simulations of \citet{cha01} for the two shorter migration time scales
in order to check the statistical consistency of the general results.
Collisions were assumed to be completely inelastic, so impacts
absorbed the mass of the smaller body into the larger one.  The
objects were assumed to be spherical, and radii were calculated using
an input density and mass.  To avoid excessive integration error for
orbits near the Sun, objects were assumed to collide with the Sun if
their heliocentric distance fell below 0.1 AU.  A timestep of 8 days
was used for the integrations, and a Burlirsch-Stoer tolerance of
$10^{-11}$.

\begin{figure}
\plotone{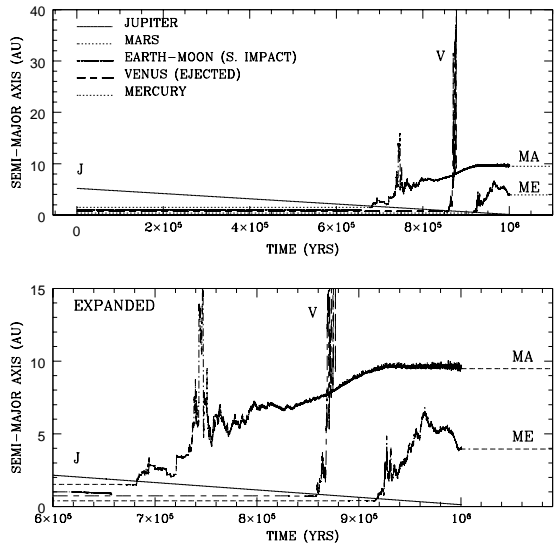}
\caption{\label{samplerun} Graph of semi-major axis vs. time for a
sample integration with a migration time scale of $1\times10^{6}\ {\rm
yr}$.  The final semi-major axis of bound planets is shown as a dotted line.
Interactions during migration result in impacts with the
central star, planet-planet impacts, ejection from the system, or
simply eccentricity enhancement. A significant fraction of scattered
planets remain bound, and dynamical friction may cause the orbits of
these planets to re-circularize.}
\end{figure}

A sample integration plot from the $10^6$ yr time scale series
of integrations using constant migration is shown in
Fig. \ref{samplerun}.  
As expected, strong dynamical evolution of the inner planets is observed.
The principal
effects of interest here are secular orbital perturbations of the
smaller planets due to evolving orbital resonances during the giant
planet's movement and close interactions with the giant planet.

As the giant planet
migrates inward, it moves through various orbital resonances with the
inner planets, and excites
large eccentricity oscillations.  The semi-major axis of the
terrestrial planet often decreases somewhat, and random scatterings of
different terrestrial planets due to their mutual interaction during
this initial excitation period may cause collisions with the Sun or
one of the other planets.  In the $10^6$ year integration, the planet
at initial Earth orbit is particularly susceptible to resonances with
the giant planet, typically locking into a 3:1 resonance and migrating
with the giant planet until it is perturbed by the Venus analog.
If a terrestrial planet survives the initial perturbation until the
giant planet is in a relatively close orbit, then direct interaction
with the giant planet will dominate the next stage of evolution.
The outcome is typically a slingshot encounter, where the terrestrial
planet is impelled outwards and its eccentricity,
semi-major axis, and orbital period increase 
sharply.\footnote{If a terrestrial planet is initially in a near-circular orbit with
semi-major axis $a_i$, it will generally only suffer a large impulsive
perturbation to its orbital elements if it comes within the ``Hill''
radius of influence, $r_H = 3\sqrt[3]{m_J/M_*}$ of the Jovian, where
$m_J$ is the mass of the Jovian and $M_*$ is the mass of the central
star \citep{glad93}.  After an impulsive perturbation the orbit has
some final eccentricity, $e$, and semi-major axis, $a_f \gg a_i $, and
necessarily $a_f(1-e) \approx a_i$.  The orbital period is now
typically 1-2 orders of magnitude longer than it was initially.  On
each return to periastron, the terrestrial planet has a renewed
opportunity to interact with the Jovian, if it is still bound to the
star. However, the probability of interaction is only $\sim r_H/2\pi
a_i \sim 0.02$ per orbit crossing, and if migration is rapid the
Jovian moves by $r_H$ in orbital radius in $\sim 10^4$ years at $a_i
\sim 1 \ AU$, so there are only a small discrete number of occasions
for interactions that could lead to ejection or collision before the
Jovian migrates far enough in that it is decoupled from the new orbit
of the terrestrial planet.  If the terrestrial planets straddle the
Habitable Zone when formed initially, then the final re-circularized
orbits of a significant fraction of surviving terrestrial planets will
be close to their initial orbits.
}
The net impulse
varies depending on the velocity difference between the terrestrial
planet and the giant planet and the distance at closest approach.  If
the planet remains in a Jovian-crossing orbit this can occur several
times, exciting the orbit dramatically and possibly leading to the
ejection of the terrestrial planet.  Planets which experience orbital
excitation due to the giant planet typically also acquire large
orbital inclinations, as opposed to planets excited by other
terrestrial planets which remain at low inclinations; similar effects
were noted by \citet{tho02} in simulations of the formation of Neptune
and Uranus.

The dynamics are chaotic, and the ultimate fate of
specific planets is highly dependent on initial conditions.  However,
once the Jovian-type planet has moved sufficiently close to the star
it effectively decouples from any remaining bound terrestrial planets
and the remaining planets settle into quasi-stable orbits.
Longer-duration simulations will be necessary to adequately evaluate
more fully the detailed long term stability of the ensemble of
surviving systems; however, in real planetary systems other effects,
including dissipative effects, and other outer planets, may change the orbital parameters of
the outer terrestrial planets before significant long term dynamical
evolution takes place.

\begin{figure}
\plotone{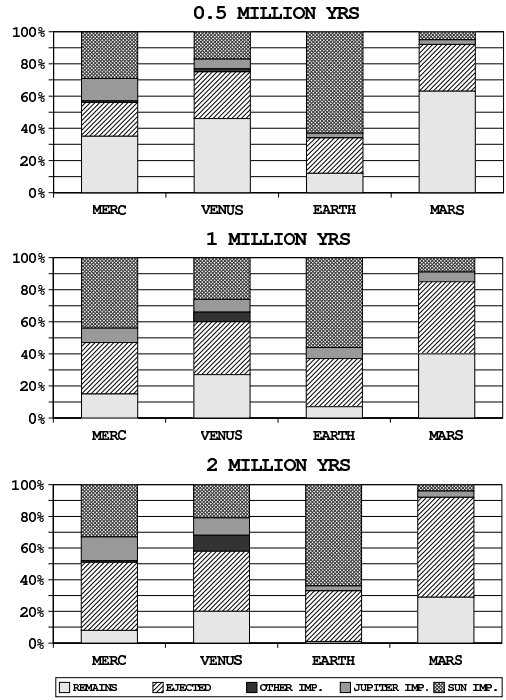}
\caption{\label{charts} Final distributions of the fate of the four
terrestrial planets for each of the three migration time scales using a
Solar System terrestrial planet formation scenario. 100 models are averaged for
each plot. The
survival probability increases as the time scale for migration
decreases.  The significant differences between the fate of the four
planets are partially the result of the different resonances with the
giant planet, and partially due to the random initial relative orbital
phases.}
\end{figure}

For the $5\times10^{5}$ year integrations, the Jovian-type planet's increased
migration speed caused fewer ejections (due to fewer crossings) and
more total remaining terrestrial planets.  For the $2\times10^{6}$
year integrations, this trend is reversed.  The percentage of planets
remaining declined from 40\% for the $5\times10^{5}$ year integrations
to 15\% for the $2\times10^{6}$ year integrations, with an overall
average of about ${1/4}$ of terrestrial planets ending up in bound
orbits outside the migrated Jovian-type planet, averaged over all
three sets of integrations for the Solar System analog using constant
migration (see Fig. \ref{charts} for a complete distribution of the
three main simulation groups).  The alternate planetary configurations
from the Chambers models gave similar results, with a maximum survival
rate as high as 35\% for one configuration and a migration time scale
of $10^6$ years.  As expected, using a variable migration model
taken from \citet{tri02}, which starts with a fast linear migration
and slows after $\sim6\times10^{5}$ years, resulted in a higher
survival rate for the Mars and Earth analogs (because of the faster
initial migration) and a lower survival rate for the Venus and Mercury
analogs (due to the slower migration rate a late times), averaging out
to a similar overall survival rate.

\begin{figure}
\plotone{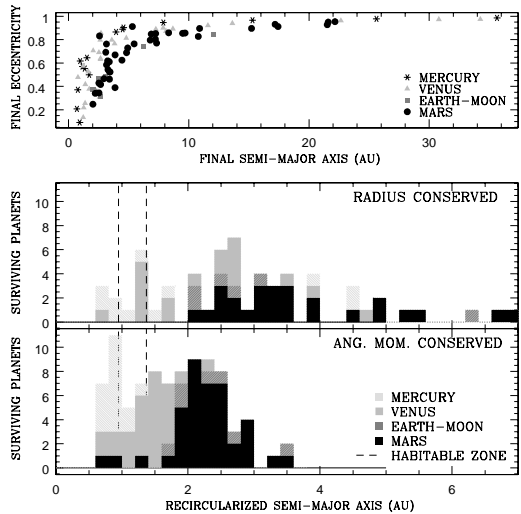}
\caption{\label{elemout} The upper plot shows semi-major axis
vs. eccentricity for all the surviving planets in the 100 integrations
with a migration time scale of $1\times10^{6}\ {\rm yr}$ and a constant
migration rate.  The correlation between semi-major axis and
eccentricity is simply due to the requirement that the final orbit
pass through the point at which the impulse on the original orbit
occurred.  A slight trend with initial planet position can be seen.
The lower plots show histograms of the theoretical final semi-major axes for
planets coming within 7 AU only, in the 100 integrations after re-circularization
due to dynamical friction has occurred.  The middle plot assumes an
upper limit where the semi-major axis is conserved after
recircularization; 7\% of the surviving planets are in the Habitable
Zone.  The lower plot assumes a lower limit of re-circularization with
conservation of angular momentum; in this case 16\% of the surviving
planets would have semi-major axes in the Habitable Zone.}
\end{figure}

Several interesting conclusions may, tentatively, be drawn from these
results.  First, despite expectations to the contrary,
the migration of a giant planet does not
eliminate terrestrial mass planets from the inner planetary system.
A significant fraction of terrestrial planets may survive
the giant planet migration process, {\it if} the embryonic inner planets
are in place before onset of type II migration.  Therefore planetary systems may
form and persist with a hot Jovian-type planet in a close orbit and
terrestrial planets at farther distances.  Future searches for
Earth-like planets need not be restricted to planetary systems like
our own.  This is particularly true if migration occurs quickly,
either due to massive protoplanetary disks or other migration effects.
This could also have ramifications for systems without a close-in
giant planet.  If migration causes a giant planet to travel inwards
and accrete onto the parent star, only terrestrial planets would be
left in the outer system. Therefore, systems in which migration of the
Jovian-type planets lead to the total destruction of the most massive
members of the system could still contain terrestrial planets which
survived the migration process and ended up in near-circular orbits in
or near the Habitable Zone.

The semi-major axes
and eccentricities of the remaining terrestrial planets span a large
range, from less than 1 AU to greater than 30 AU and from 0.1 to
almost 1.0 respectively (see Fig. \ref{elemout}). Orbital inclination is 
also generally large
at the end of migration. Similar results were found for several test
simulations using 100 asteroid-size particles, suggesting that the
interactions with the giant planet are independent of planetesimal
mass for low masses.  Current calculations suggest
{\it ab initio} planet formation cannot take place in the outer disk after
migration \citep{arm03}.  However, the presence of remaining
planetesimals and circumstellar material in the disk will lead to
strong circularization of the bound terrestrial planets due to
dynamical friction and/or interactions with the remnant gas disk
\citep{agn02,tho02}. The outer bound terrestrial planets can settle
back into near-circular co-planar orbits, even if the mass of material
in the outer disk is inadequate to form terrestrial planets from
material remaining after migration. Some incremental mass accretion may also occur.
Planets with high eccentricities will tend to recircularize
with minimal change in semi-major axis
(for discussion see \citet{tho02},
\citet{wei97} and \citet{wet89}).  Lower limits on
post-recircularization orbital size can be found by assuming
conservative re-circularization with negligible angular momentum
evolution, where the final circular orbit has semi-major axis $a_c =
a_f (1-e^2)\approx a_i(1+e)$.  
The outer disk may
provide additional drag on the outer planets, leading to migration
inwards; by inspection (Figure \ref{elemout}) modest drag will not
decrease the fraction of planets that end up in the Habitable Zone
and, if anything, will slightly increase the numbers.
Assuming a solar-type star as the
parent star and a Habitable Zone between 0.95 AU and 1.37 AU
\citep{kas93}, between 7\% and 16\% of the remaining planets in the
$1\times10^{6}\ {\rm yr}$ simulation would eventually reside within
the Habitable Zone, which implies of order 1-4\% of systems in which
migration occurred would have a terrestrial mass planet in the
Habitable Zone, assuming a pre-migration ensemble of terrestrial
planets comparable to the Solar System.  

This has very interesting implications for the existence of
terrestrial planets, and specifically habitable terrestrial planets,
in planetary systems with a migrating Jupiter-mass planet. 
Earth-like planets may persist in the Habitable Zone despite the
rapid migration of a giant planet through the Habitable
Zone during the formation phase. The critical result is that planetary
embryos assumed to form before onset of migration can survive the
migration process and persist as cores for terrestrial mass planets in
the inner system, outside the orbit of the Jovian that migrated
inwards.  The postulated re-circularization phase, post-migration,
involves interaction with primarily icy outer system objects; naively
one therefore predicts significant volatile enrichment of the planets
post-migration, possibly leading to a population of predominantly
volatile-rich terrestrial planets (see \citet{kuc03} for a discussion
of volatile-rich planets). Further research is required to explore the
detailed interaction of the planets in the post-migration phase.

In conclusion, in a simplified dynamical model of a
giant planet migrating through a stationary terrestrial planet system
terrestrial planetary embryos will {\it not} be
completely eliminated from a planetary system.  During migration,
terrestrial planets can cross the orbit of the migrating giant planet.
The scattered planets have a significant probability of remaining bound to the
central star, and may settle back into the Habitable Zone, even where
post-migration ab initio formation of terrestrial planets is not possible.

\acknowledgements
{Supported in part by
PSARC, NSF Grant PHYS 02-03046, and
contributions from the Z. Daniel Scholarship.
We thank the referee for constructive comments.
}

\end{document}